\documentclass[pre,twocolumn,aps,superscriptaddress,longbibliography]{revtex4-1}
\usepackage[left=1.5cm, right=1.5cm, top=2cm, bottom=2.0cm]{geometry}
\usepackage{graphicx}
\usepackage{dcolumn}
\usepackage{bm}

\usepackage{times}
\usepackage{amssymb}
\usepackage{lastpage}
\usepackage{float}
\usepackage{fancyhdr}
\usepackage{fnpos}
\usepackage[english]{babel}
\usepackage{array}
\usepackage{droidsans}
\usepackage{charter}
\usepackage[T1]{fontenc}
\usepackage[usenames,dvipsnames]{xcolor}
\usepackage{setspace}
\usepackage[compact]{titlesec}
\usepackage{appendix}
\usepackage{lineno}
\usepackage[colorinlistoftodos]{todonotes}
\usepackage{amsmath}
\usepackage{bm}
\usepackage{amsbsy,amsmath}
\usepackage[free-standing-units]{siunitx}
\usepackage[symbol]{footmisc}
\newcommand{\secref}[1]{Section\ \ref{#1}}

\newcommand{\figref}[1]{Fig.\ \ref{#1}}

\newcommand{\bs}[1]{\boldsymbol{#1}}
\newcommand{\mrm}[1]{\mathrm{#1}}
\usepackage[symbol]{footmisc}

\begin{document}

\preprint{APS/123-QED}

\title{Phase diagram and aggregation dynamics of a monolayer of paramagnetic colloids}

\author{An T. Pham}
\affiliation{NSF Research Triangle Materials Research Science and Engineering Center, Duke University,
Durham, NC 27708, USA}
\affiliation{Department of Mechanical Engineering and Materials Science, Duke University, Durham, NC 27708, USA}
\author{Yuan Zhuang}
\affiliation{NSF Research Triangle Materials Research Science and Engineering Center, Duke University,
Durham, NC 27708, USA}
\affiliation{Department of Chemistry, Duke University, Durham, NC 27708, USA}
\author{Paige Detwiler}
\affiliation{Department of Biomedical Engineering, Duke University, Durham, NC 27708, USA}
\author{Joshua E. S. Socolar}
\email{Corresponding author: socolar@phy.duke.edu.}
\affiliation{NSF Research Triangle Materials Research Science and Engineering Center, Duke University,
Durham, NC 27708, USA}
\affiliation{Department of Physics, Duke University, Durham, NC 27708, USA}
\author{Patrick Charbonneau}
\email{patrick.charbonneau@duke.edu}
\affiliation{NSF Research Triangle Materials Research Science and Engineering Center, Duke University,
Durham, NC 27708, USA}
\affiliation{Department of Chemistry, Duke University, Durham, NC 27708, USA}
\affiliation{Department of Physics, Duke University, Durham, NC 27708, USA}
\author{Benjamin B. Yellen}
\email{yellen@duke.edu.}
\affiliation{NSF Research Triangle Materials Research Science and Engineering Center, Duke University,
Durham, NC 27708, USA}
\affiliation{Department of Mechanical Engineering and Materials Science, Duke University, Durham, NC 27708, USA}
\affiliation{Department of Biomedical Engineering, Duke University, Durham, NC 27708, USA}

\begin{abstract}
We have developed a tunable colloidal system and a corresponding theoretical model for studying the phase behavior of particles assembling under the influence of long-range magnetic interactions. A monolayer of  paramagnetic particles is subjected to a spatially uniform magnetic field with a static perpendicular component and a rapidly rotating in-plane component.  The sign and strength of the interactions vary with the tilt angle $\theta$ of the rotating magnetic field.  For a purely in-plane field, $\theta=90\degree$, interactions are attractive and the experimental results agree well with both equilibrium and out-of-equilibrium predictions based on a two-body interaction model.  For tilt angles $50\degree\lesssim \theta\lesssim 55\degree$, the two-body interaction gives a short-range attractive and long-range repulsive (SALR) interaction, which predicts the formation of equilibrium microphases. In experiments, however, a different type of assembly is observed. Inclusion of three-body (and higher-order) terms in the model does not resolve the discrepancy.  We further characterize the anomalous regime by measuring the  time-dependent cluster size distribution.
\end{abstract}

\pacs{Valid PACS appear here}
\maketitle


\section{INTRODUCTION}

Colloidal suspensions are often used as macroscopic models of atomic systems because the cohesion energy per particle can easily be made commensurate with room temperature thermal energy, $k_{\mathrm{B}}T$, and because the length and time scales associated with the particle dynamics (micrometers and seconds, respectively) allow for facile tracking of individual particles with an optical microscope. Studies of colloidal suspensions have yielded insights into the microscopic dynamics of phase transitions such as spinodal decomposition~\cite{bailey2007spinodal}, glass formation~\cite{weeks2000three}, crystallization~\cite{van1997template} and martensitic transformations~\cite{yang2015phase,peng2015two}. In addition, these systems can easily achieve particle-scale confinement, which is more challenging to observe and control in atomic systems. Examples include  particles confined within narrow cylinders~\cite{fu2016assembly}, transport through narrow pores~\cite{huber2015soft}, and assembly near a hard wall~\cite{yang2015assembly}. The behavior of a monolayer of particles confined between two plates separated by a distance close to the particle diameter is of particular interest both for theoretical reasons, such as the stabilization of topological defects, and for engineering materials with optimized electronic, optical, or elastic properties. Although such systems have been studied extensively in computer simulations~\cite{gelb1999phase,schmidt1997phase,toxvaerd1999molecular,lichtner2012phase,lichtner2013spinodal,evans1990fluids,christenson2001confinement,alba2006effects}, realizing analogous experimental systems has been challenging.

The strength and spatial extent of particle interactions are key determinants of the phase diagram and equilibration dynamics. Tuning the interaction strength (at fixed temperature) serves as a proxy for tuning the (inverse) effective temperature. Meanwhile, tuning the spatial extent of the interactions can completely change the structure of the phase diagram and the nature of the phase transitions~\cite{ten1997enhancement}. Much attention has been paid to colloidal systems with short-ranged attractive interactions, in which the extent of the interparticle potential is smaller than roughly one quarter of a particle diameter~\cite{noro2000extended}. Examples include colloidal-polymer mixtures~\cite{tuinier2003depletion,mao1995depletion,foffi2002phase} and DNA-mediated binding~\cite{varrato2012arrested,rogers2016using,wang2015crystallization}. These systems have been especially useful for elucidating the role of metastable critical points and spinodal decomposition in physical gelation~\cite{lu2008gelation,royall2015role} and in discovering higher-order glass transitions~\cite{pham2002multiple}. By contrast, systems with attractive interactions spanning distances greater than the particle radius are expected to more closely mimic the behavior of simple liquids~\cite{hansen1990theory}, which display both a stable gas--liquid critical point and a gas--liquid--crystal triple point. Prior attempts to realize these long-range interactions in experiments have been based on colloidal-polymer mixtures with a large size ratio~\cite{teece2011ageing,zhang2013phase,sabin2016exploring} and on systems that exploit the critical Casimir effect~\cite{hertlein2008direct,faber2013controlling,edison2015critical,dang2013temperature}. The development of a colloidal model with more easily tunable long-range attraction has potential to improve the control over particle assembly and thereby open the door to the study of equilibrium phases formed from more complicated interaction potentials.

\begin{figure}[t]
  \centering
  \includegraphics[width=\columnwidth]{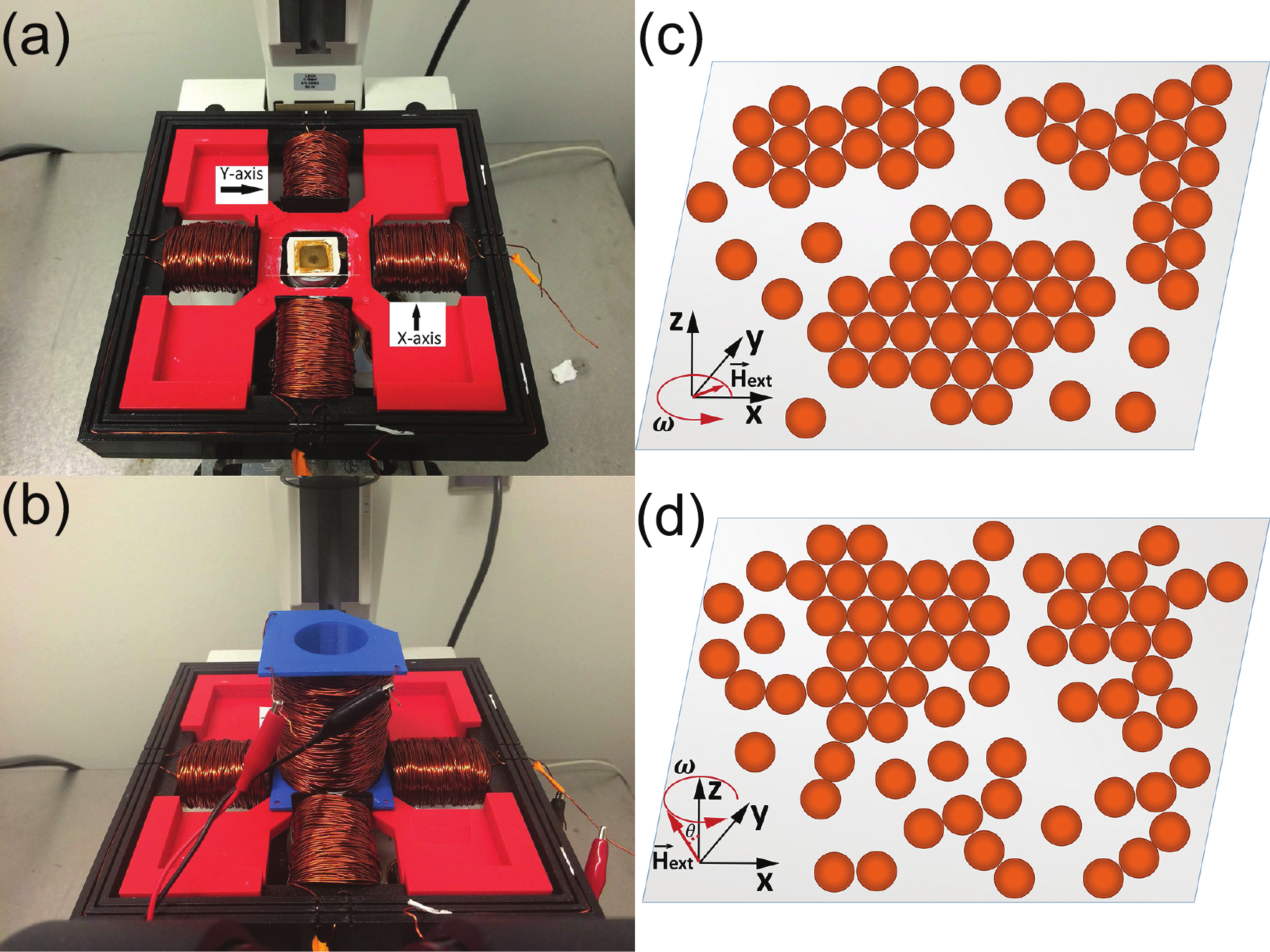}
  \caption{Experimental apparatus. (a) A colloidal suspension confined in a thin fluid film was placed at the center of a 3D printed microscopy platform that includes a pair of bi-axial solenoids for applying an in-plane magnetic rotating fields. (b) Tilted magnetic fields were achieved by adding a third solenoid on top of the microscopy platform. Illustrations of self-assembled configurations for the cases of an in-plane rotating field (c) and a conical rotating field (d) are provided.}
  \label{Experiment Apparatus}
\end{figure}

Colloidal particles with electric or magnetic dipole moments induced by an applied field exhibit interactions that decay with the inverse cube of the interparticle distance, $r^{-3}$ and can be tuned dynamically. The inherent anisotropy of these interactions, however, presents challenges for inducing bulk condensation. Although isotropic repulsive interactions can be obtained with a static external field applied perpendicularly to the monolayer~\cite{eisenmann2004anisotropic,hoffmann2006partial,fornleitner2008genetic}, net isotropic cohesion requires either a binary mixture of oppositely aligned dipoles~\cite{yang2015phase}, or the addition of a high-frequency in-plane rotating field to a monodisperse suspension~\cite{tierno2007viscoelasticity, du2013generating, mohoric2016dynamic}. The latter has the advantage that both the interaction type and its strength can be be controlled in situ (See Fig.~\ref{Experiment Apparatus}). Hence, a single experiment can both explore the system behavior at different effective temperatures and change the shape of the interaction potential on the fly. The commercial availability of colloidally stable paramagnetic particles and the straightforward setup for creating and tuning rotating magnetic fields further  facilitate the efficient study of assembly in systems with long-range attractive interactions, thus enabling the systematic characterization of phase boundaries and out-of-equilibrium assembly of colloidal systems with long-range interactions. 

The formation of condensed phases composed of paramagnetic colloids in time-dependent applied fields has been extensively studied.  A variety of putative equilibrium structures have been observed, including open-cell foams, sheets, and molecule-like clusters ~\cite{martin2003generating, martin2013driving} in three dimensions, where the singular features of conical magnetic fields with the magic opening angle of~54.7\degree have been emphasized.   Complex non-equilibrium structures, including membranes, gel-like networks, and crystallites, have also been observed in two-dimensionally confined systems, at high field strength \cite{osterman2009field,muller2014pattern, jager2012nonequilibrium, Snoswell2009, maier2016critical, Biswal2017}.  These studies suggest a need for a thorough understanding of the equilibrium phases and of the out-of-equilibrium relaxation dynamics in these systems.  

The present work shows that a quantitative match between theory and experiment can be obtained over a broad portion of the phase diagram for a 2D system, but also reveals unexplained discrepancies near the liquid-gas phase transition for conical fields near the magic opening angle. More specifically, we combine experiment and theory to calibrate and investigate the phase behavior of a two dimensionally confined monolayer of monodisperse paramagnetic particles subjected to a time-varying conical magnetic field, consisting of a rotating in-plane field and a static vertical field. By tuning the field strength and cone angle, we adjust the cohesion energy between neighboring particles from zero to a few $k_{\mathrm{B}}T$. Our experimental preparation protocol further allows us to prepare samples with a range of area fractions, $\phi=0.1-0.68$. This broad parameter space gives us access to the key features of the phase diagram, including the gas--liquid--crystal triple point and the gas--liquid spinodal regime. The carefully calibrated system is then used to study the effects of interactions obtained close to the magic tilt angle~(54.7\degree), for which the theoretical model predicts that particles should experience pair interactions that involve both short-range attraction and long-range repulsion (SALR). Such interactions are expected to produce equilibrium microphases, but have yet to be fully controlled in colloids~\cite{ciach2013origin,zhuang2016equilibrium,zhuang2016prl}. Experimentally, however, we observe a behavior more akin to Ostwald ripening than equilibrium mirophase formation. Including three-body contributions in Monte Carlo~(MC) simulations also fails to capture the observed phenomenology, and we do not observe any significant buckling of the monolayer or other macroscopic effects that are not represented in our model. Explaining the experimental behavior therefore remains an open question. 

The rest of this article is organized as follows. The experimental setup, materials and method are described in \secref{experiment_method}. \secref{computational_methods} describes the theory and numerical methods used to determine phase diagrams and simulate the colloidal dynamics.  In \secref{results}, we present the experimental results and numerical simulations of an in-plane rotating field. We also compare and contrast the types of structures obtained at different tilt angles. Finally, we summarize the results and discuss open questions in \secref{conclusions}.  

\section{EXPERIMENT METHOD}
\label{experiment_method}

The experimental system consists of magnetic spherical particles of diameter, $\sigma=$\SI{2.8}{\micro\meter} (M-270 Dynabeads\textsuperscript{\textregistered{}}, Life Technologies\texttrademark{}), with size dispersity of less than 3\%. The fluid film was prepared by placing a \SI{2.7}{\micro\liter} aliquot of the particle suspension between glass and cover slide, which was then sealed with Loctite\textsuperscript{\textregistered{}} marine epoxy. The flexibility of the coverslip induces a variation of the thickness of the fluid film. We limit our experimental observations to the thinnest regions of the sample, where particles cannot form a vertical chain, even under strong vertical fields.  Although we do observe a small number of particle pairs with overlapping images, we estimate the fluid thickness to be less than $1.5\sigma$ in all of the reported experiments.  We note also that the magnetic interaction energy is commensurate with the thermal energy, but is an order magnitude lower than the gravitational energy associated with raising a bead a distance $\sigma$. Thus the monolayer is also  maintained by gravitational confinement. The particle area fraction was kept within a range of $\phi\approx0.1-0.68$. To reduce the non-specific adhesion between particles and substrates, we grew a \SI{40}{\nano\metre} thick poly(oligo(ethylene glycol) methyl ether methacrylate) (POEGMA) polymer brush on both the glass slide and the coverslip using the surface-initiated atom-transfer radical polymerization (SI-ATRP) approach reported previously~\cite{hucknall2009simple}.

The sample was placed on a 3D printed platform that has orthogonal pairs of solenoids oriented to produce a uniform rotating magnetic field (see \figref{Experiment Apparatus}a). To obtain a conical magnetic field, a third solenoid was placed on top of the sample~(see \figref{Experiment Apparatus}b) thus generating a static magnetic field along the $z$-direction. The magnetic field strength in each direction was calibrated using a Lakeshore\texttrademark{} 410 hand-held Gaussmeter with a resolution of $\SI{0.1}{Oe}$. Geomagnetic fluctuations in our laboratory have mean square variation of less than $H\approx\SI{0.05}{Oe}$ during experimental time frames, which is more than an order of magnitude smaller than the weakest magnetic field that we applied.  We calibrated the applied field at frequencies in the range of $\SIrange{50}{1000}{Hz}$ by measuring the induced voltages in an oscilloscope. We obtained optimal results for an applied frequency of $f=\SI{200}{Hz}$, which was used for the remainder of our experiments.  The strength and frequency of the rotating field were controlled with Labview software (National Instruments\texttrademark{}, Version 2014, Austin, Texas). 

Previous studies of individual Dynabeads M-270 have reported the existence of permanent magnetic moments in addition to the paramagnetic response \cite{janssen2009controlled}. Such moments would be expected to lead to chain formation in weak fields \cite{jager2011pattern}.  We note, however, that we do not observe any indication of chain formation in our apparatus in this regime, suggesting that the permanent moments of the particles are rather small. It has also been shown that the magnetic moment of a bead in a rotating field lags the field by a small amount at high frequencies.  Were this effect significant, we would expect a macroscopic anisotropy to develop because the reversal of the angular velocity of the field every cycle would not allow for a full azimuthal average of the induced moments. Yet no such anisotropy was observed.

An inverted Leica\texttrademark{} DMI6000B microscope (LEICA, Bannockburn, IL) was used to record the self-assembly process through a 40x air-immersion objective. The microscope was capable of automated focusing, and images were captured at a rate of two frames per minute with a Qimaging Micropublisher\texttrademark{} 5.0 RTV Camera with resolution of 2560 x 1920 pixels (Qimaging, Surrey, Canada). A custom code was written in MATLAB~(Mathworks\textsuperscript{\textcopyright}, Version
2014, Natick, MA) for image processing and centroid identification of particles in every frame. The details of this procedure can be found in a prior study~\cite{Pham_2016}.

The flexibility of the experimental system allows us to generate a wide range of system conditions, enabling the assembly of various colloidal phases. Particles were initially  suspended homogeneously, then the external magnetic fields were suddenly increased to $H\approx \SIrange{0.6}{1.4}{Oe}$ and held constant for the rest of the experiment.  This protocol mimics a rapid cooling of the system. Because the apparatus was maintained at room temperature, $T^{\mathrm{exp}} \approx 298 \kelvin$, experiments only changed the magnetic field strength, i.e., the effective temperature, of the system.


\section{THEORY AND COMPUTATIONAL METHODS}
\label{computational_methods}
This section first describes a theoretical model of the experimental system. The model assumes that an individual bead can be treated as a paramagnetic point dipole whose moment instantly aligns with the local field at the center of the bead. Next, we describe the simulation approach used to determine the phase behavior of that model and elucidate its dynamics. 

\begin{figure}[ht]
 \centering  
 \includegraphics[width=0.8\columnwidth]{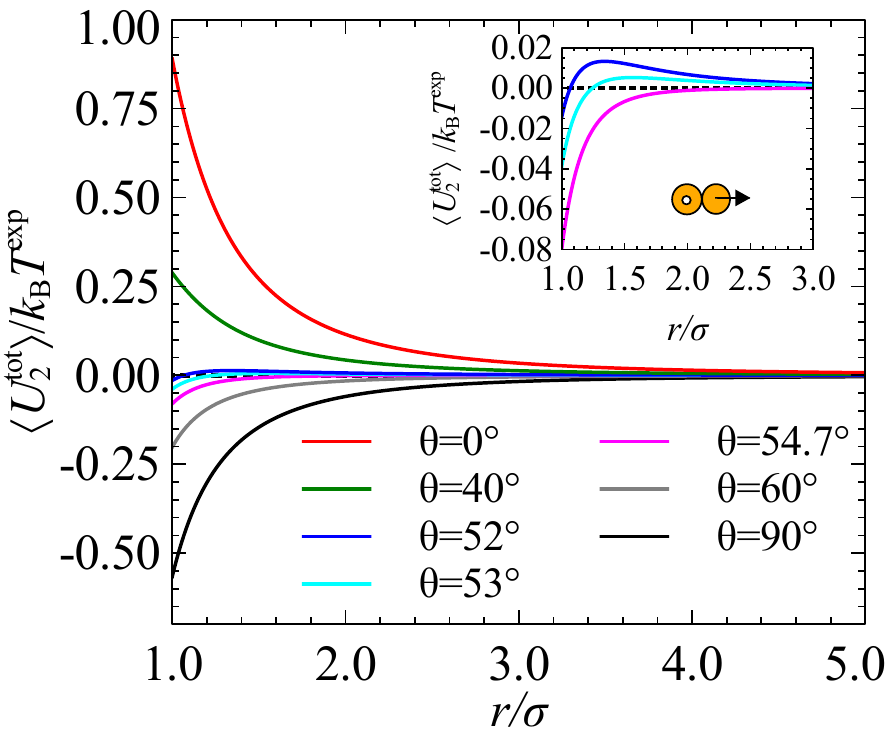} 
\caption{Pair interaction potential as a function of distance at fixed field strength, $H=\SI{1}{Oe}$, particle magnetic susceptibility, $\chi=1$, and $T^{\mathrm{exp}}\approx 298 \kelvin$, for various tilt angles $\theta$. The inset zooms in on the angular regime $52\degree\lesssim\theta<\theta_\mathrm{c}\approx54.7\degree$, over which a SALR interaction is observed at this level of approximation.}
  \label{pair-potential}
\end{figure}

\begin{figure*}[ht]
 \centering  
 \includegraphics[width=0.8\textwidth]{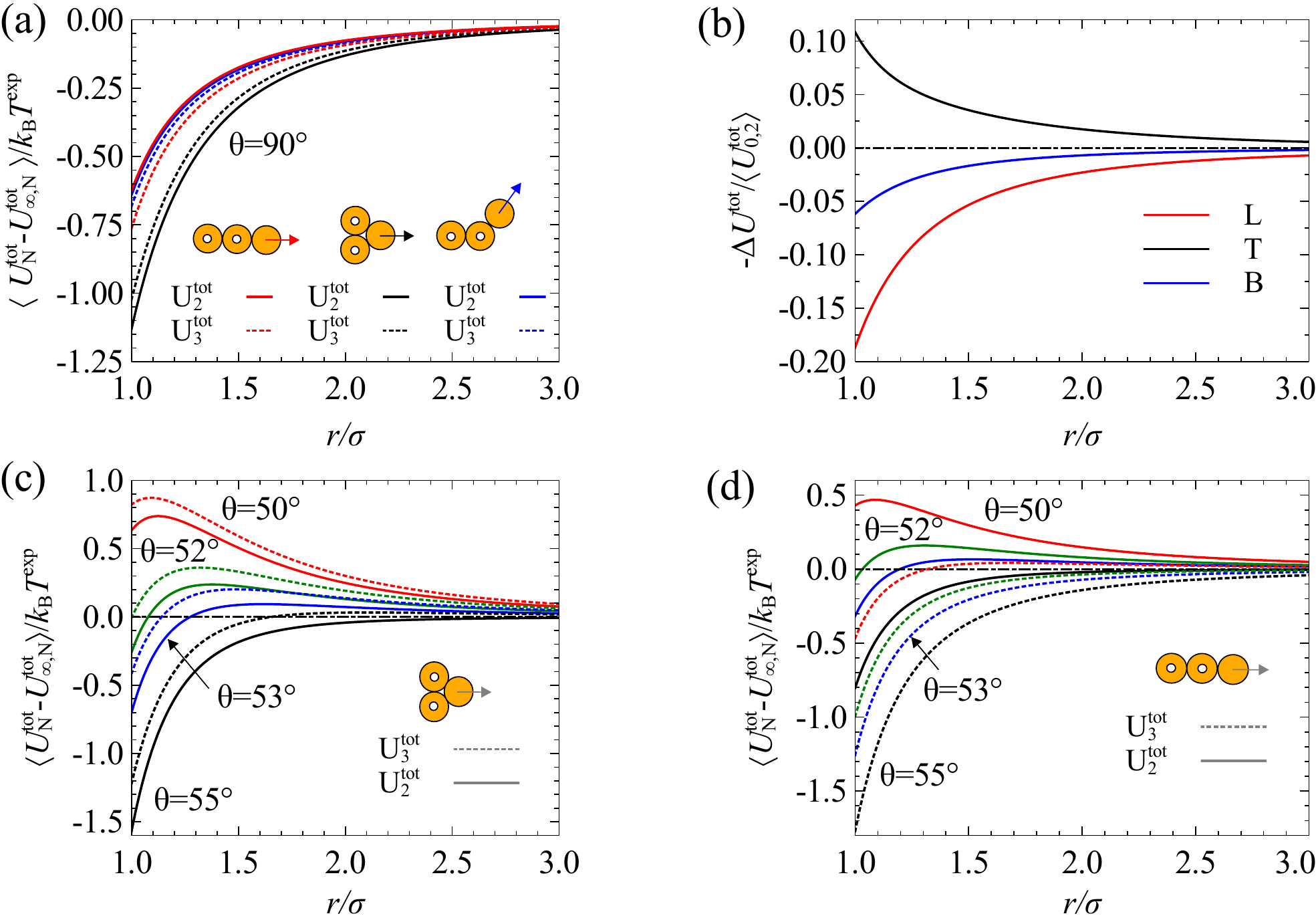} 
\caption{(a) Interaction energies for linear chain ($\mrm{L}$), triangular ($\mrm{T}$), and bent chain ($\mrm{B}$) configurations with the two particles marked by white dots held fixed. 
Energies are obtained by two-body (solid lines) and three-body (dotted lines) calculations at fixed field strength, $H\approx \SI{1}{Oe}$, and particle magnetic susceptibility, $\chi=1$, for $\theta=90\degree$. (b) Ratio of the energy discrepancy between the two- and three-body calculations, normalized by the two-body $r=\sigma$ contact energy, $U_{0,2}$. (c,d) Interaction energy for isosceles and linear chain configurations from two- (solid lines) and three-body (dotted lines) calculations for $H \approx \SI{3}{Oe}$ and tilt angles in the SALR range. The field is chosen such that the interaction energies are on the order of $k_\mathrm{B}T^{\mathrm{exp}}$ at $\sigma < r < 2\sigma$.  Note the sign difference between the three-body corrections in (c) and (d).}
  \label{three_particles_potentials}
\end{figure*}

\subsection{Modeling the interaction energies}
\label{modeling_interactions}

The rotating external field used in experiments is described as
\begin{equation}\bs{H}(\theta,\phi) =H\,\big(\sin\theta\cos\phi,\,\,\sin\theta\sin\phi,\,\,\cos\theta\big),
\end{equation}
where $H$ is the magnetic field strength, $\theta$ is the tilt angle of the field with respect to the vertical axis, which is taken to be normal to the monolayer plane, and $\phi(t) = \Omega(t) t$ with 
\begin{equation}
\Omega(t) = \left\lbrace \begin{array}{ll} 
+\omega &\mathrm{for\ }  \lfloor \omega t/(2\pi) \rfloor \mathrm{\ even} \\
-\omega & \mathrm{for\ } \lfloor \omega t/(2\pi) \rfloor \mathrm{\ odd}
\end{array}\right. \,.
\end{equation}
The instantaneous dipole moment of particle $i$ induced by the external magnetic field and the field of neighboring particles is then
\begin{equation}
  \bs{m}_{i}(t) =
  \chi{V}
 \left(\bs{H}+\sum_{j=1,j\neq{i}}^{N}{\frac{3\left(\bs{m}_{j}\cdot \bs{\hat{r}}_{ij}\right)\bs{\hat{r}}_{ij} -\bs{m}_{j}}{4\pi\left|r_{ij}\right|^3}}\right),
 \label{self-consistent equation}
\end{equation}
where $\chi$ and $V$ are the susceptibility and volume of particle $i$, respectively, $r_{ij}$ is the distance between particles $i$ and $j$, and $\bs{\hat{r}}_{ij}$ is the unit vector pointing from  $i$ to $j$. This linear set of equation can be condensed as
\begin{equation}
  \bs{M} =\chi V(\tilde{\bs{H}} + \alpha_0\bs{D} \cdot \bs{M})\,,
 \label{self-consistent matrix equation}
\end{equation}
where 
\begin{equation}
\bs{M} \equiv [m^{x}_{1},m^{y}_{1},m^{z}_{1},\dots, m^{x}_{N},m^{y}_{N}, m^{z}_{N}]^\intercal
\end{equation}
is a vector having the $3N$ associated with the $N$ particle dipole moments,  
\begin{equation}
\tilde{\bs{H}} \equiv [H^{x},H^{y},H^{z}, \dots,H^{x},H^{y},H^{z}]^\intercal
\end{equation}
is a vector having the $3N$ components of the external field acting on each particle, and $\alpha_0\bs{D}$ is the $3N\times 3N$ matrix representing the contribution at each dipole of the field due to the other dipoles. Taking $\alpha_0 = 1/(4\pi\sigma^3)$, the portion of $\bs{D}$ representing the field of particle $2$ at the position of particle $1$ is the $3\times 3$ block with dimensionless elements: 
\begin{equation}
  D_{ij}^{(12)} = \left( \frac{3\bs{\hat{r}}_{12,i}\bs{\hat{r}}_{12,j}-\delta_{ij}}{|r_{12}/\sigma|^3}\right).
 \label{matrix elements}
\end{equation}
The solution for the dipole moments is then
\begin{equation}
  \bs{M} =  \chi V[\bs{I}-\alpha\bs{D}]^{-1} \cdot \tilde{\bs{H}}\,,
 \label{self-consistent solution}
\end{equation}
where $\bs{I}$ is the identity matrix and $\alpha\equiv \chi V \alpha_0$ is a dimensionless coupling strength, which for $\chi\approx 1.0$  is of order $1/24$.

The instantaneous interaction energy of the system can be written as~\cite{kwaadgras2014self,kwaadgras2014orientation,troppenz2015nematic}
\begin{align}
  U(\theta,t) & = -\frac{\mu_{\mrm{0}}}{2}[ \bs{M}\cdot\tilde{\bs{H}}(\theta,t) - N\chi V H^2 ] \\ 
  \ & = -\frac{\mu_{\mrm{0}}}{2} \sum_{i=1}^{N}[\bs{m}_{i}(t) \cdot \bs{H}(\theta,t) - \chi V H^2],
\label{N_body}
\end{align}
where $\mu_{\mrm{0}}$ is the vacuum permeability and the $H^2$ terms remove the constant contributions associated with infinitely separated particles.

For a system of $N$ particles, calculating the full $N$-body energy requires inverting a $3N\times 3N$ matrix in Eq.~(\ref{self-consistent solution}), which is an operation that becomes prohibitively time consuming if done repeatedly for large $N$.  The result, however, can be approximated by a series of sums over subsets of $n$ particles.  
Let $u_n(1,\ldots,n)$ be the energy calculated from Eqs.~\eqref{self-consistent solution} and~\eqref{N_body} for $n$ particles at positions $\bs{r}_1,\ldots \bs{r}_n$, considered in isolation, and let $U_n$ be the total energy obtained by summing $u_n$ over all combinations of $n$ particles.
The sum over two-body contributions alone,
\begin{equation}
U_2 = \sum_{i<j}u_2(i,j), 
\end{equation}
gives a result correct to order $\alpha$, and the correct result to order $\alpha^2$ is 
\begin{equation}
U_3 = \bigg(\sum_{i<j<k} u_3(i,j,k)\bigg) - (N-3)U_2.
\label{eqn:u3tot}
\end{equation}
Subtracting the $U_2$ term in the latter expression is required because a given pairwise interaction is counted $N-2$ times in the sum over triplets of particles.  A useful alternative expression is 
\begin{equation}
U_3 = \bigg(\sum_{i<j} u_2(i,j)\bigg) + \bigg(\sum_{i<j<k} u'_3(i,j,k)\bigg),
\label{eqn:U3tot2}
\end{equation}
where
\begin{equation}
 u'_3(i,j,k) \equiv u_3(i,j,k) - u_2(i,j)-u_2(j,k)-u_2(k,i).
 \label{U'_3}
\end{equation}
Because for our system $\alpha\approx 1/24$, it suffices to consider at most terms to order $\alpha^2$. For greater accuracy, straightforward combinatorics gives
\begin{equation}
U_4 = \bigg(\sum_{i<j<k<\ell}\!\!\!\! u_4\bigg) -\, (N-4)U_3\, -\, \frac{(N-4)(N-3)}{2}U_2,
\end{equation}
correct to order $\alpha^3$, and similar expressions can be obtained for higher orders.

In the experimental system, effective isotropic interactions between particles are achieved by rotating the magnetic field about the vertical axis with a period much shorter than the timescale for particles to diffuse their own diameter, $\sigma$. In this regime, the effective interaction energy for a given tilt angle $\theta$ is given by 
\begin{equation}
  \begin{aligned}
  \langle U(\theta)\rangle = \frac{1}{2\pi} \int_{0}^{2\pi}\!\! U(\theta,\phi)\, \mrm{d}\phi\,,
\label{average pair-potential equation}
  \end{aligned}
\end{equation}
where $\langle\cdot\rangle$ indicates a time average, or equivalently, an azimuthal average.

The $6\times 6$ matrix of Eq.~\eqref{self-consistent solution} with $n=2$ can be inverted analytically, and its time average yields 
\begin{equation}\label{eq:u_exp}
\left\langle u_{2}(i,j;\theta) \right\rangle = \frac{\mu_0 \beta_{ij} \chi V H^2}{1+\beta_{ij}}\left(1+\frac{3\big(\cos(2\theta)-1\big)}{4(1-2\beta_{ij})}\right),
\end{equation}
where $\beta_{ij} = \alpha \sigma^3/ r_{ij}^{3}$. The resulting pair interaction potential with distance $r$ is shown in \figref{pair-potential} for different field tilt angles.  For $\theta>\theta_\mathrm{c}=(1/2)\cos^{-1}(-1/3)\approx54.7\degree$, the interaction is purely attractive, while for $\theta < \theta_c$ the energy peaks at a finite separation distance, producing a repulsive force beyond that point.  For $\theta\lesssim 52\degree$, the interaction is purely repulsive for all $r>\sigma$.  For tilt angles in the range $52\degree\lesssim\theta<\theta_\mathrm{c}$, however,  a SALR potential is obtained, as detailed in the inset of \figref{pair-potential}. In dense suspensions, the two-body approximation is affected by the presence of many other surrounding particles, leading to an unexpected anisotropic interaction. Hence the higher-order approximations are likely essential in this regime.

Three-body (order $\alpha^2$) contributions to the energy are expected to be important when unbinding particles produces a very small change to $U_2$.  A rough approximation for when this happens is obtained by assuming a uniform distribution of particles beyond $\sigma$, and finding $\theta_0$ such that 
\begin{equation}
\int_{\sigma}^{\infty} \langle u_2(r;\theta_0)\rangle\,r\,{\mrm d}r = 0.
\end{equation}
A numerical evaluation yields $\theta_0 \simeq 53.9\degree$.
Figures~\ref{three_particles_potentials}a and~\ref{three_particles_potentials}b compare the two-body to the three-body interaction energy for a single set of three particles in an in-plane rotating field. In each of these three cases, the energy is plotted as one of the particles is moved away while the other two are held fixed, and $r$ denotes the distance between the moving particle and the nearer of the other two particles. Figure~\ref{three_particles_potentials}a compares the self-consistent two-body,  $\langle U_2 \rangle$, and  three-body, $\langle U_3 \rangle$,  energy for  a linear chain ($\mrm{L}$), a bent chain ($\mrm{B}$), and an isosceles triangle ($\mrm{T}$).  
Figure~\ref{three_particles_potentials}b shows the difference~($\Delta{U}$) between the corresponding energies. 

For an in-plane rotating magnetic field, $\theta=90\degree$, the energy difference  between two- and three-body calculations is $10$-$20\%$ for compact configurations ($r/\sigma=1$ in \figref{three_particles_potentials}b), which is  comparable to the experimental measurement error. The minimum energy configuration is the equilateral triangle, as shown in~\figref{three_particles_potentials}a, and the three-body correction is positive. For the linear and bent chains, the three-body correction has the opposite sign, as is clear in~\figref{three_particles_potentials}b. In both cases, the minimal energy configuration is the close-packed one, which is indeed the local arrangement that is most commonly observed in experiments (see \figref{phase_snapshot} below).

In the SALR regime, $52\degree\lesssim\theta<\theta_\mathrm{c}$, the situation is more complex. For the isosceles configuration shown in  \figref{three_particles_potentials}c, the two-body approximation underestimates the energy, whereas for straight chains (\figref{three_particles_potentials}d) it overestimates it. (See also Ref.~\cite{martin2003generating}) In this regime the three-body contribution is therefore qualitatively significant. 
\subsection{Simulation and phase diagram determination}
The colloidal experiments are done at fixed (room) temperature, $T^{\mrm{exp}}=298K$, with different applied field strengths. Because the interaction energy is proportional to the square of the external field strength (see Eq.~\eqref{N_body}), we can also define an effective temperature 
\begin{equation}
T = \frac{k_\mrm{B} T^{\mathrm{exp}}}{\mu_0 V H^2},
\label{TandH}
\end{equation}
and then rescale the interaction energy such that $\mu_0 V H^2 = 1$ in Eqs.~\eqref{N_body} and~\eqref{eq:u_exp}, as is standard in simulations~\cite{frenkel2002understanding}. 
The two-body approximation to the energy used in simulations is then a reduced form of Eq.~\eqref{eq:u_exp} and truncated at $10\sigma$ for computational efficiency.
Because the truncation error is less than $1\%$ of the total system energy tail corrections are not performed~\cite{frenkel2002understanding}. 
For MC simulations with $\alpha \approx 1/24$, this treatment gives a reasonable approximation of the full energy for tilt angles far from the SALR regime. 

\begin{figure*}[t] 
 \centering  
 \includegraphics[width=0.8\textwidth]{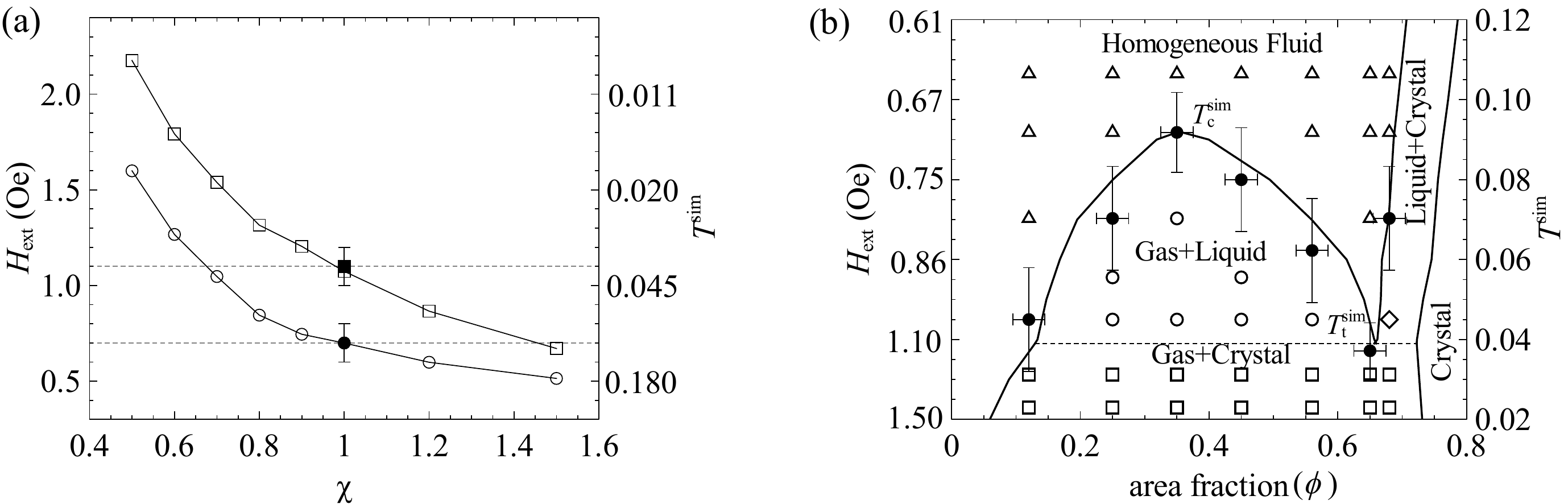} 
 \caption{(a) The critical ($\circ$) and triple ($\square$) point temperatures from MC phase diagrams as a function of magnetic susceptibility, $\chi$, compared with the observed experimental values denoted by a solid dot ($\bullet$) and a solid square ($\blacksquare$), respectively. The agreement between experiments and simulations identifies the particle susceptibility, $\chi\approx 1.0(1)$. (b) The simulation phase diagram for the two-body interaction potential with $\chi=1.0$ is compared with the experimental results. Solid dots ($\bullet$) denote the experimental phase boundaries with the error bars on the applied field strength and area fraction. Triangles ($\vartriangle$) denote the homogeneous fluid region; circles ($\circ$) the gas-liquid coexistence region; squares ($\square$) the gas-crystal coexistence region; and diamond ($\diamond$) the liquid-crystal coexistence region. MC coexistence results at different temperatures are interpolated in order to obtain the solid black lines. Note that in this system the critical temperature $T_{\mathrm{c}} \approx 0.092(2)$ corresponds to a critical magnetic field strength $H_{\mathrm{c}} \approx {0.7(1)} \SI{}{Oe}$ at $\phi \approx 0.35$, while the triple point $T_{\mathrm{t}} \approx 0.039$ corresponds to a magnetic field strength $H_{\mathrm{t}} \approx {1.1(1)} \SI{}{Oe}$ at $\phi \approx 0.64(3)$.}
  \label{phase_diagram}
\end{figure*}

In the SALR regime the two-body contribution is small, hence three-body contributions are relatively more important~\cite{martin2003generating}.
Because $U_3$ is prohibitively expensive to calculate, we implement a cutoff on the range of the three-body term by including in Eq.~\eqref{eqn:U3tot2} only those triplets for which the maximum distance between two particles is 3 particle diameters, $3\sigma$.
As analytic expressions are not available for $u_3$, we generate a look-up table for $u'_3$~(Eq.~\eqref{U'_3}) for configurations of three particles in which no particle pair is beyond the distance cutoff. 
Algorithmically, $u'_3$ is initially computed in distance increments of $0.01\sigma$, and then interpolated for a specific configuration from the look-up table. Contributions from this term are then added to $U_2$.

Phase diagrams for the simulation model are obtained using Monte Carlo-based free energy methods. These approaches rely on thermodynamic integration~(TI) schemes varying either pressure $P$ and temperature, or the spring constant of an Einstein crystal for the Frenkel-Ladd  approach~\cite{frenkel2002understanding}. 

In the gas and liquid regimes, constant $NPT$ MC simulations are performed for different pressures $P$ along an isotherm above the critical temperature $T_\mathrm{c}$. An approximation of the liquid equation of state is then obtained by interpolating the simulation results with cubic splines. Isothermal TI of this equation from the low-density ideal gas provides the fluid free energy as a function of density at that temperature. The results are then used as starting points for different isobaric TIs, which give  the free energy of lower-temperature systems, near the  gas--liquid coexistence regime.
The gas--liquid phase boundary for a given temperature is then determined by a common tangent construction from the free energy results on both side of the coexistence regime. 
The critical point is extracted by fitting the coexistence curve to the $2$D Ising universality scaling
\begin{equation}
\label{eq:crit}
  \rho_{\pm} = \rho_{\mathrm{c}}+2C_2\left|1-\frac{T}{T_\mathrm{c}}\right|\pm \frac{1}{2}B_0\left|1-\frac{T}{T_\mathrm{c}} \right|^{\beta_{\mathrm{c}}},
\end{equation}
with the corresponding critical exponent $\beta_{\mathrm{c}}=1/8$.
Note that the Gibbs ensemble Monte Carlo scheme, which often provides a computationally more direct way of delimiting gas--liquid coexistence~\cite{frenkel2002understanding}, is here inefficient. The very low surface tension between the two disordered phases indeed leads to facile interface formation within a single simulation box, even for $T$ well below $T_\mathrm{c}$. 

For the crystal phase, an Einstein crystal of non-interacting particles tightly pinned by an ideal spring to a perfect lattice was used as reference. During the subsequent Frenkel--Ladd TI, the spring constant is gradually reduced to zero. These simulations were done at an area fraction, $\phi=0.63$--$0.86$, chosen such that the triangular crystal is stable for the given $\chi$ and finite system size. From these reference points, integration over $P$ and $T$ provided the free energy of crystals at nearby state points.
Note that because of the relatively strong attraction between particles much larger systems would be needed for the quasi-long-range nature of the two-dimensional order to play a quantitatively noticeable role and no hint of an intermediate hexatic phase was detected in simulations. Hence, the contribution of these effects to the liquid stability regime, which is here our main interest, would be small.

Each simulated state point ran for $10^6$ MC steps, each step consisting of $N$ local particle displacements, tuned so the acceptance ratios are between $40\%$ and $60\%$. For the constant $NPT$ simulations of the gas and liquid regimes, $N=1000$ and one logarithmic volume change per MC step was included on average. For the constant $NVT$ simulations of the crystal phase, $N=864$ was used and the Frenkel--Ladd integration was done using $20$-point Gaussian--Lobatto quadratures over logarithmically spread spring constants, from $\lambda_{\mrm{max}}=1000$ down to $\lambda=0$.

\subsection{Structural observables}
\label{Structural_Observables}

\begin{figure*}[t]
  \centering
  \includegraphics[width=\textwidth]{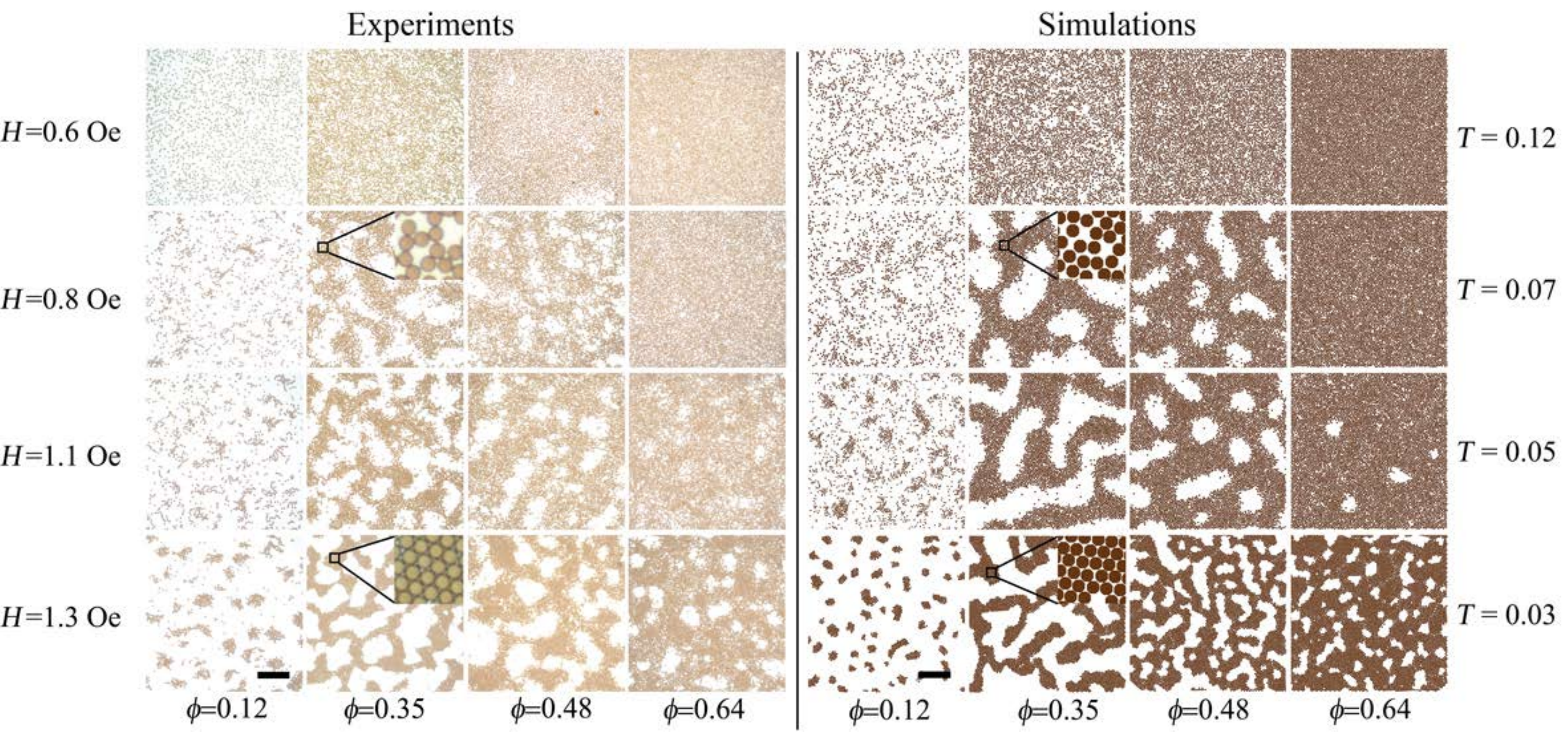}
  \caption{Particle configurations at $\phi\approx0.12-0.64$ under magnetic strengths fields in the range of $H \approx\SIrange{0.6}{1.3}{Oe}$ ($T \approx 0.12-0.03$). These snapshots are taken 60 minutes after quenching to the desired field in experiments (left column), which roughly corresponds to 6000 MC steps in simulations (right column), as determined in Section~\ref{Structural_Observables}. Each image covers  $235\micro\meter \times 235 \micro\meter$, and the scale bar is $50 \micro\meter$ long. Insets detail the local liquid and crystalline order on either side of the triple point temperature.  Note that in experiments the space between nearly touching particles appears filled due to optical effects.}
  \label{phase_snapshot}
\end{figure*}

To gain insight into the structure and dynamics of the gas--liquid phase separation,  experimental systems were quenched to fields above the critical field, while simulated systems were instantaneously quenched  to temperatures below the critical point. The former are obtained under the same conditions as in Section.~\ref{experiment_method}, while the latter were initially prepared for $N=8000$ equilibrated at $T=5.0\gg T_\mathrm{c}$ and instantaneously quenched using constant $NVT$ MC simulations with only local particle displacements. Comparing the two approaches at a same state point provides a conversion factor for matching simulation to experiment dynamics. We find that 100 MC steps correspond to roughly 1 minute in experiment.

At different time points of both the experiment and simulation trajectories, we calculate the structure factor,
\begin{equation}
S(\left|\bs{k}\right|) = \frac{1}{N}\left\{\left[\sum_i \cos(\bs{k}\cdot\bs{r}_i)\right]^2 + \left[\sum_i\sin(\bs{k}\cdot\bs{r}_i)\right]^2\right\},
\label{Eq_s_k}
\end{equation}
where $\bs{r}_{i}$ is coordinate of particle $i$, $\bs{k}=\left(\frac{2\pi m}{l_x},\frac{2\pi n}{l_y}\right)$ with integers $m$ and $n$. The position $k_\mrm{m}$ and intensity~$S(k_\mrm{m})$ of the low-wavevector peak of the structure factor are obtained by performing a local Gaussian fit to the $S(k)$ results. 
Simulation results are averaged over 20 independent trajectories, while experimental results are averaged over three independent realizations.  

For experiments in tilted fields, we also consider the mean radius size of the coarsening clusters undergoing crystal nucleation and growth. Particles within crystal-like structures can be identified based on local bond-order parameters. This requires first identifying the set of nearest-neighbors for each particle from a Voronoi tessellation of the system. As in Ref.~\cite{Pham_2016}, we then evaluate the $6$-fold bond-order parameter for each particle
\begin{equation}
  \Phi_{6}^{i} \equiv
  \begin{cases}
    0, & n_\mathrm{n} < 3 \\
    \displaystyle
    \frac{1}{n_\mathrm{n}} \sum_{j=1}^{n_\mathrm{n}} \exp(6\mrm{i} \varphi_{ji}),
    & n_\mathrm{n} \geq 3
  \end{cases},
\end{equation}
where $\varphi_{ji}$ is the angle between the $x$ axis and the vector from particle $i$ to particle $j$. Particles with fewer than three nearest-neighbors, i.e., $n_{n}<3$, are removed from the calculation in order to prevent chain-like structures to bias the cluster analysis. Particle $i$ and its nearest-neighbors are then classified as being part of a same cluster if the real part of their bond-orientational order correlation, $\Phi_{6}^{i} \Phi_{6}^{\ast j}$, exceeds 0.1. Such a low threshold is chosen in order to accurately identify all particles within clusters, including less ordered ones at the cluster surface. 
The process completes when each particle in the field of view is either associated with a cluster or labeled as chain-like.     

Defining the center of mass of each cluster $m$
\begin{equation}\label{Centroid}
  \bm{C}_{m}=\frac{\sum_{i=1}^{N_m}\bm{r}_{i}}{N_m},
\end{equation}
where $N_m$ is the number of particles in $m$  and $\bm{r_i}$ denotes particle $i$ position, we can determine the cluster radius
\begin{equation}\label{Cluster_radius}  R^{m}_{C}=\sqrt{\frac{\sum_{i=1}^{N_m} (\bm{r}_{i}-\bm{C}_{m})^2}{N_m}}.
\end{equation}
Averaging over all clusters $m\in M$ gives the average cluster radius
\begin{equation}
  \langle R_{C} \rangle =  \frac{1}{M} \sum_{m=1}^{M} R^{m}_{C}.
\end{equation}

\section{RESULTS}
\label{results}
Our study consists of two sets of experiments and simulations.  First, we consider the coarsening and phase behavior under an in-plane rotating field ($\theta = 90\degree$), for which the interparticle forces are attractive at all distances beyond contact. In this case, we obtain a quantitative match between experiment and simulation.  We construct a complete experimental phase diagram and show that both the critical and the triple point can be identified. By fitting the location of these two points, we determine the particle magnetic susceptibility, $\chi$, and find good agreement between the resulting simulation phase diagram and experiments.  We also find a good match between observed experimental and simulated dynamics following quenches into the gas-liquid coexistence regime after identifying the experimental timescale associated with a MC step.

Second, we consider the coarsening and phase behavior under magnetic fields tilted at various angles away from the vertical. Here, we observe a rich set of behaviors that includes surprising structural features for $50\degree\lesssim\theta\lesssim55\degree$, which is near the regime in which the two-body interparticle potential is SALR.  In this regime, straightforward extension of the simulated potential produces qualitatively different configurations from the experimental data.  To characterize this regime more fully, we present additional experimental analysis of cluster formation and growth.   

\begin{figure}[tbh]
 \centering 
 \includegraphics[width=\columnwidth]{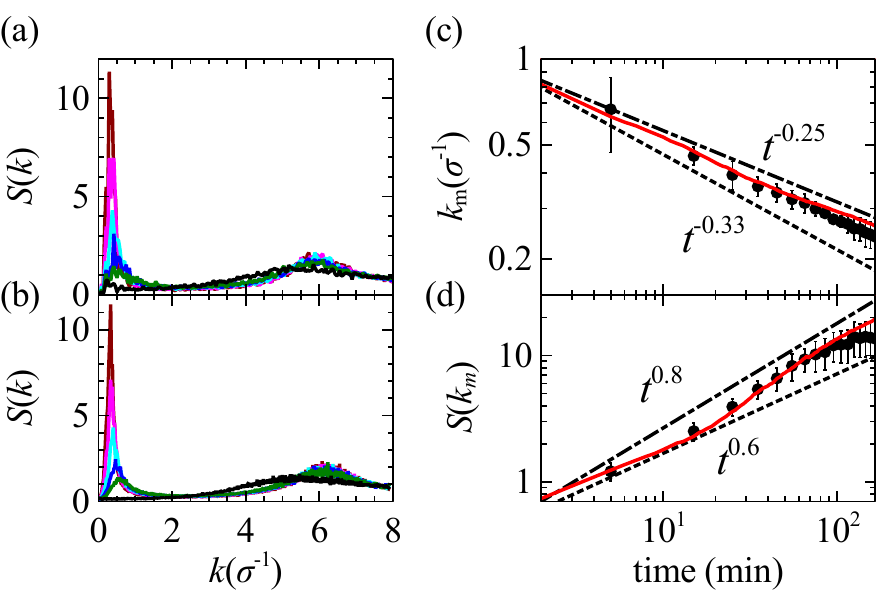} 
 \caption{Structure factor, $S(k)$, at different times ($t=0, 5, 10, 20, 40, 60$ minutes after quenching in black, green, blue, cyan, magenta, and dark red lines, respectively) in (a) experiments and (b) MC simulations at corresponding state points, $H \approx \SI{0.8}{Oe}$ ($T\approx 0.07$), and area fraction,~$\phi\approx 0.35$. (c) Dynamical scaling of the low wavevector peak positions,~$k_{m}$~(filled circle), averaged over three different experimental runs. Error bars denote 95\% confidence interval on the mean. Simulation results~(red line) are averaged over $20$ different trials. The power laws indicate diffusion-limited kinetics in the coarsening of the colloid-rich domains with scaling exponents of ~$\gamma=0.33$~(dot) and~$\gamma=0.25$~(dash-dot), respectively. (d) Time evolution of~$S(k_{m})$ in experiments~(filled circle) and simulations (red line). The dotted and dash-dotted line are power laws with exponents~$\zeta=0.6$ and~$\zeta=0.8$, respectively.}
  \label{s(k)}
\end{figure}

\subsection{In-plane rotating fields}
\label{In-plane rotating}


Particles exposed to an in-plane rotating magnetic field ($\theta=90\degree$) experience isotropic interactions that are purely attractive within the plane of the monolayer and penalize out-of-plane motion. In order to determine the magnetic susceptibility of the particles, $\chi$, we first experimentally identify the critical and triple points. The former is the lowest field strength at which the gas and liquid phases coexist, and the latter is the lowest field strength at which crystallites appear. Because both are here visually identified, their determination is achieved with the same precision as that used for experimentally selecting state points, i.e., $\pm \SI{0.1}{Oe}$ for the field strength and $\pm2.5\%$ for the particle density. 
By comparing these two characteristic fields and densities to the simulation results for the two-body approximation at various $\chi$ (\figref{phase_diagram}a), we find that $\chi=1.0(1)$ provides the best fit. Interestingly, this result is consistent with the vendor-provided value, $\chi\approx0.96$~\cite{byrom2013magnetic}, as well as with a value independently determined in prior work, $\chi\approx1.2(2)$~\cite{van2013accurate}. We thus have a reasonably high degree of confidence in our characterization of this material property. We note, however, that reported values for the same particles range from $\chi \approx 0.17$ to $\chi\approx 1.45$~\cite{helseth2004microscopic,van2015biological,helseth2007paramagnetic}. 

Setting $\chi=1$, we align the experimental field strength to the simulated temperature in the two-body model~(See Eq.~(\ref{TandH})) Figure~\ref{phase_diagram}b shows the experimental phase behavior superimposed on the gas--liquid, gas--crystal and liquid--crystal phase boundaries from MC simulations. The experimental error bars encompass these boundaries, indicating that the model accurately reproduces the experimental behavior. Because the experimental error is comparable to that of the pairwise approximation at this $\theta$ (which is roughly 10\%, see~\figref{three_particles_potentials}b), this agreement further suggests that no systematic error was made in determining the experimental coexistence line. We note, however, that  reliably obtaining high-density configurations is experimentally challenging, hence the liquid--crystal coexistence boundary could not be precisely determined. 

\begin{figure*}[ht]
 \centering 
  \includegraphics[width=\textwidth]{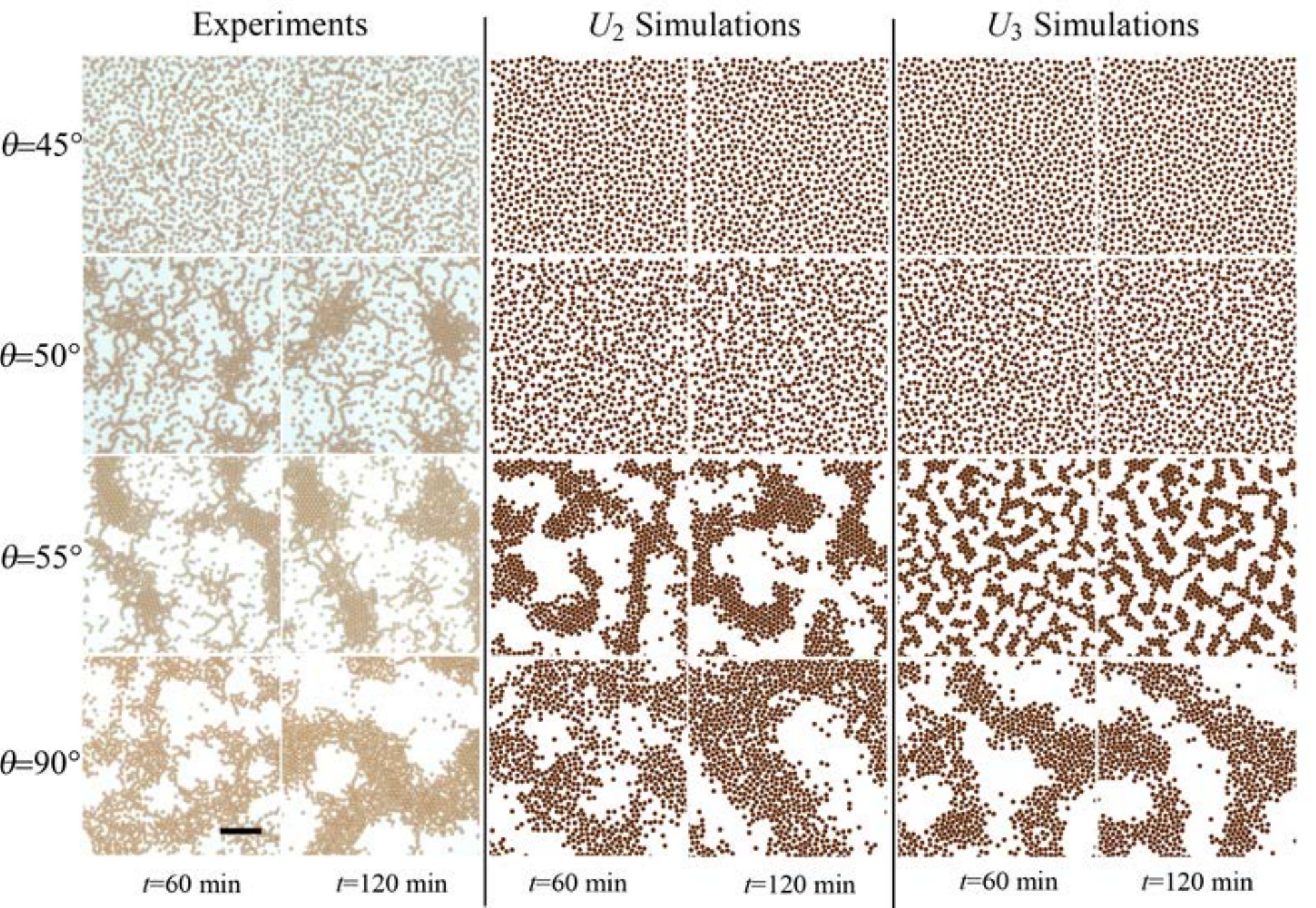} 
 \caption{Evolution of particle assembly in experiments and simulations at different field tilt angles~(rows). The magnetic field strength is held constant at $H \approx \SI{3}{Oe}$ ($T\approx0.005$) for both experiments~(left column) and simulations with $\chi=1$~(middle and right columns) at $45\degree\leq\theta\leq55\degree$, while the in-plane field strength is kept at $H \approx \SI{0.8}{Oe}$ ($T \approx0.07$). Simulations are taken at similar tilt angles, field strength and particle area fraction~($\phi\approx0.35$) for both two-~(middle) and three-body~(right) models. Each experiment frame covers $150\micro\meter\times150\micro\meter$, and the scale bar is $30\micro\meter$ long. The dynamical scaling factor determined in Section~\ref{In-plane rotating} is used to establish a correspondence between time and MC steps.}
  \label{tilt_angle}
\end{figure*}

The snapshots in~\figref{phase_snapshot} illustrate the correspondence between the behavior of the pairwise model and experiments. Observed particle configurations~(left) are quite similar to those obtained in simulations~(right). In weak magnetic fields ($H \approx \SI{0.6}{Oe}$, $T \approx 0.12$), particles are uniformly distributed for all densities ($\phi \approx 0.12-0.64$), as is expected in a homogeneous fluid (see the top row of \figref{phase_snapshot}). As the field strength increases (or temperature is lowered), the system phase separates into colloid-rich (liquid phase) and colloid-poor (gas phase) regions that coexist for a range of field strengths  ($H \approx \SIrange{0.8}{1.0}{Oe}$~, $T \approx 0.05-0.07$) and area fractions ($\phi \approx 0.35-0.48$)~(see supplemental movie S$1$~\cite{supplemental_material}). For densities below or above this regime the system remains homogeneous (second row of \figref{phase_snapshot}). Finally, as the magnetic field is increased beyond $H \approx \SI{1.1}{Oe}$ ($T \approx 0.04$), gas--crystal coexistence can be observed~(see bottom two row of \figref{phase_snapshot} and supplemental movie S$2$~\cite{supplemental_material}).  

The remarkable agreement between experiment and theory for the equilibrium behavior suggests that our colloidal system offers a high degree of control. Such control is fairly common for colloidal suspensions with short-range attractive interactions, but in that case the gas--liquid--crystal triple point does not exist and the gas--liquid critical point is metastable~\cite{binder2014perspective}. Studies of the few other systems in which reasonably long-range attraction have been obtained have encountered other difficulties. First, although critical Casimir forces can be used to control gas-liquid and liquid-solid coexistence~\cite{faber2013controlling}, the colloid densities needed to reach the critical and triple points have thus far remained inaccessible. Second, although polymer-colloid mixtures with size ratios close to unity have provided a well-characterized critical point~\cite{zhang2006experimental,teece2011ageing,sabin2016exploring}, unambiguously identifying their triple point has remained challenging~\cite{tuinier2008phase,teece2011ageing}. The display of both stable critical and triple points coupled with the ability to dynamically adjust the effective temperature by tuning the external magnetic field and thereby observe a large portion of the full phase diagram in a single experiment makes our system especially promising.

To further validate the theoretical description, we consider the dynamics of gas--liquid phase separation. More precisely, we investigate how a homogeneous system coarsens upon turning on the rotating in-plane magnetic field. Because we expect this process to be akin to spinodal decomposition, the structure factor, defined in Eq.~\eqref{Eq_s_k}, is used to quantify its time course in both experiment~(\figref{s(k)}a) and simulation~(\figref{s(k)}b). Once again, the two approaches give remarkably similar results. At early times, a single peak is observed at a spatial frequency slightly below $k\approx 2\pi/\sigma$, which is consistent with the mean particle separation distance of the homogeneous  initial configuration. Gradually with time, this peak shifts to slightly higher wavevectors, which signals that particles gradually become more densely packed. 

Although no low wavelength peak is initially present in the homogeneous fluid, one quickly develops as phase separation begins. Physically, the low-$k$ peak captures the typical separation between colloid-rich regions as the system coarsens. 
As expected for a system undergoing spinodal decomposition, the time evolution of the wavenumber of this peak~(see \figref{s(k)}c) is consistent with a power-law scaling,~$k_{m}\propto t^{-\gamma}$, at least over the time decade accessible in experiment. The resulting exponent, $0.25<\gamma<0.33$, lies within a range consistent with the kinetic coarsening described by the Cahn-Hilliard equation~\cite{barrat2003basic},
and is consistent with other simulation and experimental results for 2D systems~\cite{amar1988monte, rogers1989numerical,fratzl1994kinetics,blunt2007coerced}. The peak magnitude, $S(k_{m})$, also seems to increase as a power law (\figref{s(k)}d) with coarsening exponent, $\zeta=0.7(1)$. Interestingly, this observation is consistent with the diffusion-limited mechanism for spinodal decomposition, which predicts $\zeta>2\gamma$~\cite{furukawa1984dynamics}. 

\subsection{Conical rotating field}
\label{conical field}

Having calibrated the system as described in Section~\ref{In-plane rotating}, we now consider its behavior in a conical field.  Figure~\ref{tilt_angle} provides snapshots at different tilt angles (rows) and times (columns).  The experimental (left) and MC simulation results for the two-body model (middle)  qualitatively agree at high and low $\theta$, outside of the SALR regime at $52\degree\lesssim\theta<\theta_\mathrm{c}$.
Within the putative SALR regime, however, marked differences are observed. Simulations suggest that a stable bicontinuous morphology consistent with a disordered microphase regime should develop~\cite{zhuang2016equilibrium}, while experiments present a network of clusters connected by thin filaments.

\begin{figure}[ht]
 \centering 
  \includegraphics[width=\columnwidth] {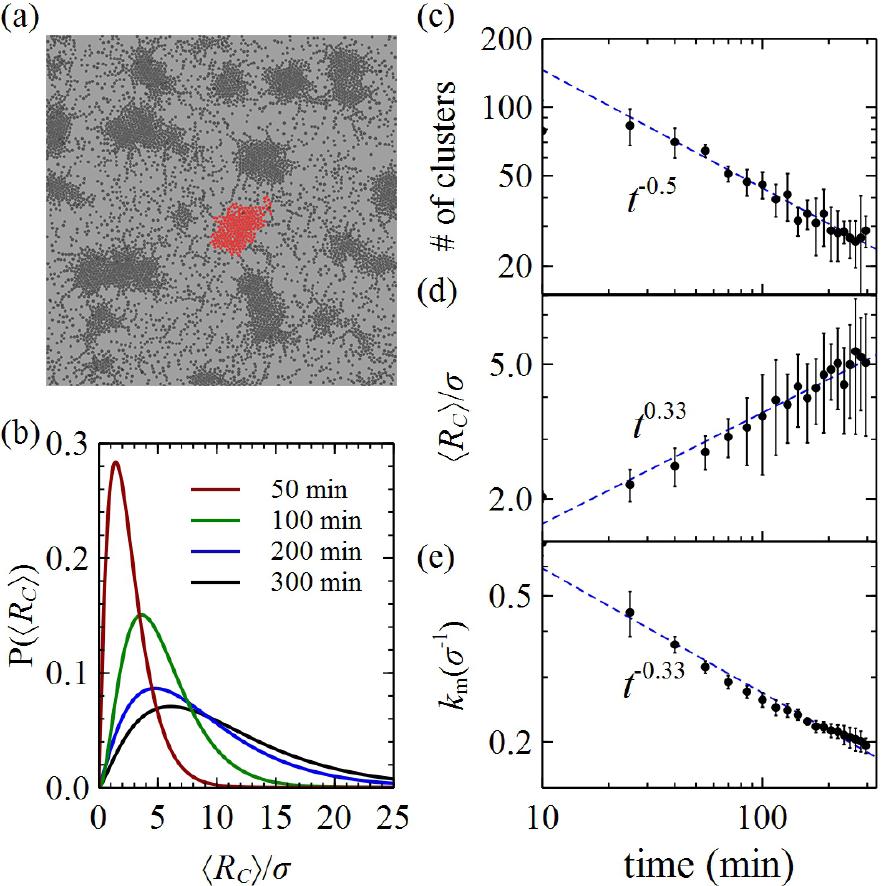} 
 \caption{(a) Droplet morphology in experiments at $\phi\approx0.35$, $\theta\approx53\degree$, and  $H \approx \SI{3}{Oe}$ after $200$ min. The highlighted domain shows one cluster identified as described in \secref{Structural_Observables}. (b) The cluster size  distribution at different experimental times. (c-e) Time dependence of the number of clusters, the mean cluster radius, and the low-$k$ peak position. Results are averaged over three different trials at fixed field strength and particle area fraction.  Dashed blue lines are guides to the eye showing power laws with exponents $-0.5$, $0.33$, and $-0.33$, respectively.}
  \label{tilt_angle_dynamic}
\end{figure}

Because in the SALR regime two-particle interactions are particularly weak compared to $k_\mathrm{B}T$, one might expect the rich morphology observed in this regime to emerge when three-body or higher order
contributions are included in the simulation model. As seen in \secref{modeling_interactions}, three-body interactions are indeed proportionally much stronger in the SALR range than at larger or smaller
tilt angles. Furthermore, in this regime the three-body interactions favor short linear chains over small close-packed clusters \cite{osterman2009field}. Large clusters remain favored over linear chains, however, suggesting the possibility of a complex balance between cluster and chain formation in equilibrium in the temperature (or field strength) regime of interest here.  For $\alpha \approx 1/24$, explicit computations of the approximate energy $U_3$ and the fully self-consistent energy $U$ for chains and clusters of up to 24 particles show discrepancies of less than 1\%.

To assess whether three-body interactions are sufficient to explain the experimental observations, we performed simulations that include them. As expected, at high and low fields the three-body contribution has but a limited impact. In the SALR regime, by contrast, three-body effects suppress the formation of large clusters and lead to the formation of elongated clusters a few particles in width, as seen in the $U_{3}$ column of Fig.~\ref{tilt_angle} at $\theta=55^{\degree}$.  Nevertheless, the match between the resulting structures and the experimental observations is not as strong as outside the SALR regime.

The experimental dynamics in the SALR regime is quite distinct from what is seen in simulations or in the spinodal regime (Section~\ref{In-plane rotating}). 
We observe a process of classical nucleation and growth at early stages.  At longer times, however, several long-lived clusters are seen to evaporate and subsequently redeposit their particles on larger clusters~(See Experimental Movie S$3$~\cite{supplemental_material}). Based on this observation and results showed in~\figref{tilt_angle_dynamic}, we interpret this coarsening mechanism as Ostwald ripening~\cite{baldan2002review}.
Although this dynamical process has been observed in a variety of systems, such as unbalanced binary liquid mixtures~\cite{li2012phase,hore2010dissipative} and late diffusion-limited spinodal decomposition~\cite{huse1986corrections}, its physical origin in the current system remains unclear.

To quantitatively characterize cluster growth, we consider the time evolution of various structural observables \secref{Structural_Observables}. \figref{tilt_angle_dynamic}b shows the changing experimental cluster size distribution as time goes. The most probable peak shifts to larger radii and broadens, consistent with theoretical and experimental studies of Ostwald ripening~\cite{conti2002phase,blunt2007coerced}. Further validations of the mechanism are provided by the decrease in the number of clusters and the growth of the mean cluster size, as shown in \figref{tilt_angle_dynamic}c and \figref{tilt_angle_dynamic}d, respectively. The growth of the mean cluster size is consistent with a power-law scaling with an exponent ${0.33}$, as predicted for Ostwald ripening, and the low-wavevector peak at~$k_{m}$ decays with a similar form, consistent with the classical Lifshitz-Slyozov theory~\cite{komura2012dynamics}.
\section{CONCLUSIONS}
\label{conclusions}
We have developed a system for studying the phase behavior of colloidal particles with long-range attraction that allows us to straightforwardly tune the interaction strength, thus enabling the direct observation of the critical and triple points. The latter, in particular, is rarely, if ever, detected in colloidal experiments.

The ability to change the form of the interaction form from attractive to repulsive by changing the cone angle of the rotating magnetic field further gives access to a rich set of other phases, including a SALR regime that is not fully understood.  Our theoretical predictions based on a system of interacting, inducible point dipoles suggest the possible formation of microphases for fields near the magic angle, when the pairwise interaction is SALR. Experiments, however, display a completely different behavior in this regime ~($\theta\approx 50-55\degree$). A mix of clusters and filaments forms and the assembly mechanism is akin to Ostwald ripening. 

In attempting to identify the physical origin of this unexpected behavior, we were able to rule out several possible sources. We believe that many-body interactions with $n>3$ are unlikely to be important because the coupling strength $\alpha$ is small for our system. This effect is also unlikely to be caused by irreversible aggregations, such as particles getting pinned to the glass slide or to each other. Though we observe particles buckling at $45\degree$, this rarely occurred in the SALR regime. One possibility is that the effect is caused by a breakdown of the point dipole model. The error introduced by the point dipole approximation, though not large, is indeed most significant near the magic angle~\cite{kwaadgras2014self}. It is also possible that other interactions not explicitly considered here, such as DLVO and steric effects, may play a role. Finally, we cannot rule out the possibility that  permanent dipole moments of the particles or complex susceptibilities at high frequencies are responsible for observed clustering in the SALR regime. Incorporating these effects into the Monte Carlo simulations is a possible yet challenging direction for further investigation. In any case, for now the question remains open for consideration by the broader soft matter community.
\section*{ACKNOWLEDGMENT}
We thank Ashutosh Chilkoti for advice and laboratory support in preparing the POEGMA coated substrates. This work was supported by the National Science Foundation Research Triangle Materials Research Science and
Engineering Center (DMR-1121107).
\bibliographystyle{apsrev4-1}
\bibliography{PRE}
\end{document}